\documentclass[fleqn,10pt]{wlscirep}
\pdfoutput=1
\usepackage{bm}
\usepackage{blindtext}
\usepackage{booktabs}
\usepackage{pdflscape}
\usepackage{appendix}
\usepackage{algorithm, algpseudocode}
\usepackage{epsfig}
\newcommand{\ra}[1]{\renewcommand{\arraystretch}{#1}}
\newcommand{\dsoriginal}[1]{{\it Original Dataset}}
\newcommand{\dssmall}[1]{{\it Small Dataset}}
\newcommand{\dslarge}[1]{{\it Large Dataset}}
\newcommand{\preoriginal}[1]{{\it Original}}
\newcommand{\presmall}[1]{{\it Small}}
\newcommand{\prelarge}[1]{{\it Large}}

\title{Irreproducibility; Nothing is More Predictable\\
{\it \small In these matters the only certainty is that nothing is certain} {\small Pliny the Elder, AD23-79}}

\author[1]{ David, Kohn}
\author[2, 5]{Nick Glozier}
\author[2, 5]{Ian B. Hickie}
\author[3, 6]{Hugh Durrant-Whyte}
\author[3,4]{Sally Cripps\thanks{Corresponding author.}}

\affil[1]{\footnotesize Sydney Informatics Hub, University of Sydney, Sydney, 2006, Australia}
\affil[2]{\footnotesize Brain and Mind Centre, University of Sydney, Sydney, 2006, Australia}
\affil[3]{\footnotesize Centre of Translational Data Science, University of Sydney, Sydney, 2006, Australia}
\affil[4]{\footnotesize School of Mathematics and Statistics, University of Sydney, Sydney, 2006, Australia}
\affil[5]{\footnotesize Faculty of Medicine, University of Sydney, Sydney, 2006, Australia}
\affil[6]{\footnotesize Faculty of Engineering, University of Sydney, Sydney, 2006, Australia}

\date{}

\begin{abstract}
The increasing ease of data capture and storage has led to a corresponding increase in the choice of data, the type of analysis performed on that data, and the complexity of the analysis performed. The main contribution of this paper is to show that the subjective choice of data and analysis methodology substantially impacts the identification of factors and outcomes of observational studies. This subjective variability of inference is at the heart of recent discussions around irreproducibility in scientific research. To demonstrate this subjective variability, data is taken from an existing study, where interest centres on understanding the factors associated with a young adult's propensity to fall into the category of `not in employment, education or training' (NEET). A fully probabilistic analysis is performed, set in a Bayesian framework and implemented using Reversible Jump Markov chain Monte Carlo (RJMCMC).  The results show that different techniques lead to different inference but that models consisting of different factors often have the same predictive performance, whether the analysis is frequentist or Bayesian, making inference problematic.  We demonstrate how the use of prior distributions in Bayesian techniques can be used to as a tool for assessing a factor's importance.
\end{abstract}

\begin{document}

\flushbottom
\maketitle

\thispagestyle{empty}

\noindent

\section*{Introduction}

Much has been written about the lack of reproducibility of scientific research in general, \cite{Ioannidis2005, Gelman2014}, and in epidemiology and public health in particular, \cite{Peng2006}.  
There are many explanations put forward for this lack of reproducibility, which, broadly speaking, can be grouped into two categories. 
The first category is behavioural, such as the tendency to submit, publish and cite positive findings more often than negative ones. 
Psychology and psychiatry are possibly the worst offending disciplines in this category, \cite{Fanelli2010, Fanelli2012}.  
The second category relates to the choice, quality and understanding of the statistical analysis. 
This category includes the confounding prevalent in observational studies, \cite{Goodman2017}, the failure to account for model uncertainty and multiplicities, \cite{Scott2010}, and the misuse and variability of the p-value, \cite{Nuzzo2014, Halsey2015}.  

Of course, the two categories are related: the bias for publishing positive results affects the choice of data used and the type of analysis performed.  
This is what \cite{Simmons2011} and  \cite{Gelman2014} refer to as `degrees of researcher freedoms'. 

In this paper we aim to show just how difficult the task of identifying factors which affect an outcome can be, when using observational data, and the consequent lack of reproducibility.  
To do this we perform a Bayesian analysis using data from a previous study \cite{ODea2014}, where interest centres around  the  identification of factors that most influence an individual's propensity to fall into the category of `not in employment, education or training' (NEET) \cite{OECD2012}.  This group of disengaged young people form a key political concern in many high income societies \cite{Mascherini2012}. 
If a young person misses out on the key transition from education into adult roles they face a lifetime of social, economic and health disadvantage. 
 
 One key factor which influences a young adult's propensity to become, and stay, disengaged appears to be poor mental health \cite{Scott2013}. 
In Australia, a large programme of open access services for young adults aged 12- 25, predominantly aimed at mental ill-health and associated problems, has been established and  showed a much higher proportion of those seeking help at such services are NEET, than the youth population in general, \cite{ODea2014}.   
In this cross sectional study NEET status in young people with mental health problems was influenced by age, sex, depression, substance use and criminal charges, but not education or cultural background as suggested by previous work, \cite{Bynner2002}. 
Longterm NEET status was associated with chronic depression, or poorer cognitive function, but these inferences depended upon how the data were assessed, \cite{Odea2016}.  
 
The study by \cite{ODea2014} used a frequentist approach and provided a single final model, based on stepwise regression, limiting the number of potential explanatory factors according to some criteria. 
This is common practice in analytical epidemiology, where the goal is to identify a set of explanatory factors, often associated with each other, and derive causal inferences, acknowledging study design limitations and biases. 
In contrast to this frequentist approach, the Bayesian approach proposed in this paper, produces an ensemble of models, each model weighted by its posterior probability,  acknowledging that these posterior probabilities depend on what was used as the initial certainties, or `prior' probabilities.   In this fashion a Bayesian approach accounts for model uncertainty referred to by  \cite{Scott2010}.

The different results we present for this one cohort study, highlight how  those wishing to prioritize or target interventions may be misled by the development of single models as opposed to approaches which use an ensemble of models, such as the Bayesian approach presented in this paper.

The dataset in the  \cite{ODea2014} study is of a very modest size by today's standards, both in terms of the number of people and the number of predictive factors; it has observations on approximately 600 individuals and 50 possible explanatory factors. Yet we show that, even with a relatively small dataset, the conclusions  depend upon the type of analysis performed.  
This is not surprising when one considers that the number of possible factor combinations that one can construct from 50 factors is $2^{50}$, which is approximately the number of insects on earth. 

This paper makes four contributions. 
\begin{enumerate}
\item
It demonstrates that the choices available to researchers in data processing and analysis, can lead to very different narratives for the factors underpinning the outcome, in this case youth disengagement.
\item
It highlights the difficulty of identifying a `final model' of important factors and combination of factors that best predicts an outcome and shows that the predictive performance, as measured by the receiver operating curve (ROC) area under curve (AUC) metric, can be very similar for models of different factor combinations. 
\item
It presents a Bayesian approach to quantifying model uncertainty, and compares the predictive performance of a Bayesian model averaging approach with the model selection approach of \cite{ODea2014}. 
\item
It investigates how the necessity of explicitly stating a prior probability for a factor's inclusion, can be used as a tool for assessing a factor's importance.
\end{enumerate}
\section*{Methods}
The methods sections is divided into three parts; a model and priors section which describes the Bayesian model and prior assumptions; a prediction and estimation section, where we describe how the parameters of the model are estimated and how predictive distributions are obtained; a data section where we give an overview of the data and provide references to a more detailed description.
\subsection*{Model and Priors}
\label{sec:models_and_priors}
Suppose we have observations on $n$ individuals' outcome (NEET status), 
$\bm y=\left(y_1,\ldots,y_n\right)$, 
where $y_i=1$ if  individual $i$ is classified as NEET and $y_i=0$ otherwise. 
We have corresponding measurements on $P$ possible explanatory factors (often called co-variates in the statistical literature) for each of these individuals,  contained in 
$X=(\bm x_1',\ldots, \bm x_n')'$, 
where, $'$, denotes the transpose and $\bm x_i = \left(1,x_{i1},\ldots,x_{iP} \right)$, is the $i^{th}$ row of $X$ and contains the $P$ factor measurements for individual $i$. Note that the $1$ at the beginning of each row is for the intercept (or "offset" in the machine learning literature). For example, if  we have $P=2$ factors, age and depression score, and if individual $1$ is 22 years old with a depression score of 0.5, then $\bm x_1=(1, 22, 0.5)$.

A common way to model the dependence between $y$ and $x$ is to use a generalized linear model (GLM),

\begin{equation}
\Pr\left(y_i=1|\bm x_i\right)=g\left(\bm x_i\bm\beta\right),
\label{eq_bin_reg}
\end{equation}
where the notation $\Pr(y_i=1 |\bm x_i)$, can be translated as {\it  the probability that individual $i$ is classified as NEET, given the factor measurements, $\bm x_i$, for that individual}. In equation \eqref{eq_bin_reg},  $\bm \beta=\left(\beta_{0},\ldots,\beta_{p}\right)'$ is the $(P+1) \times 1$ vector of regression coefficients and $g$ is some link function which maps a value on the real number line to the interval $[0,1]$.  A common choice for $g$ in this setting is a cumulative distribution function (CDF) such as a standard normal CDF (probit) or logistic (logit) CDF.
This article uses the standard normal CDF as the link function, which we denote by $\Phi(.)$, so that the data augmentation method of \cite{Albert1993} can be implemented to facilitate the sampling scheme used to estimate the regression coefficients and quantify uncertainty. 

Given the regression coefficients and the factors, the likelihood function is the product of the probability mass functions for $n$ independent Bernoulli random variables and is given by,

\begin{equation*}
\Pr(\bm y|X,\bm\beta)=\prod_{i=1}^n \Pr\left(y_i=1|\bm x_i\right)^{y_i}\times \left(1-\Pr\left(y_i=1|\bm x_i\right)\right)^{1-y_i}\\
\label{eq_like1}
\end{equation*}
where $\Pr\left(y_i=1|\bm x_i\right)=\Phi(\bm x_i\bm\beta)$.

To fully specify the model we place priors on those parameters needed to evaluate the likelihood, namely the regression coefficients. 
We wish to place priors on these regression coefficients to allow for the possibility that some factors on which we have measurements, do not have an  impact on the future NEET status of an individual.  

Specifically we introduce an indicator vector $\bm\gamma=(\gamma_1,\ldots,\gamma_P)$, where  $\gamma_k=1$ if factor $k$ is in the model and $\gamma_k=0$ otherwise.   Note that each possible value of the vector, $\bm \gamma$, represents a combination of factors, or model. 
For example, if there are 4 possible factors and if $\bm\gamma=(1,0,0,1)$, then this means that factors one and four are in the model, while factors two and three are not. 

We follow \cite{Smith96},  \cite{Holmes2006} and \cite{Lamnisos2009}, and many others, and allow the prior on the regression coefficients, $\bm \beta$, to depend upon this indicator vector as well as the factor measurements in $X$.
Let  $A_1=\{k\ge1:\gamma_k=1\}$ be the set of indices corresponding to those factors which are included in the model, and  $A_0=\{k\ge1:\gamma_k=0\}$  be the set of indices corresponding to those factors which are not included. 
Then our on $\bm\beta$ prior is
\begin{eqnarray*}
\bm \beta_{A_1}|\bm \gamma, X & \sim  &N\left(0,\Sigma_{A_1}\right) \\
\bm \beta_{A_0}|\bm \gamma, X & \sim & \delta(0)
\end{eqnarray*}
The symbol ``$|$'' means {\it conditional on} while the symbol ``$\sim$" means {\it is distributed as}. For example the notation $\bm \beta_{A_1}|\bm\gamma, X \sim N\left(0,\Sigma_{A_1}\right)$ can be translated as {\it Given $X$ and $\bm\gamma$,  the regression co-efficients have a normal distribution with mean zero and covariance matrix $\Sigma_{A_1}$}. In this paper $\Sigma_{A_1}$ is  pre-specifed  and depends upon those factors corresponding to the set $A_1$.  The notation $\beta_{A_0}|X,\bm \gamma \sim \delta(0)$ can be translated as {\it the regression coefficients which are not included in the model, are identically zero.} 

To specify a prior on $\bm \gamma$, we define $w_k = \Pr \left( \gamma_{k} = 1 \right)$  to be the prior probability that the $k^{th}$ factor is in the model.
There are a number of choices for $\bm w=(w_1,\ldots, w_P)$. 
For example, we could assume  that the probability that one factor is included in the model is independent, {\it a priori}, of the probability that another factor is included. 
In this case the prior for $\bm\gamma$ is $\Pr(\bm\gamma)=\prod_{k=1}^P w_k^{\gamma_k}(1-w_k)^{1-\gamma_k}$. 
We could then choose  the value of $w_k$ to correspond to some prior belief by eliciting expert opinion and have  different  values of $w_k$ for different factors and if we believe the probability that factor $k$ is in the model depends upon the presence of factor $j$ then we could induce dependency between $w_k$ and $w_j$, by altering the prior on $\bm \gamma$.  

If we don't wish to assume a particular value for $w_k$, then we could replace the value of $w_k$ with a prior distribution, for example $w_k \sim U[0,1]$, which means that $w_k$ is equally likely to take any value between 0 and 1. 

For this paper we take the subjective prior that $w_j$ is independent of $w_k$, and to account for multiplicity we follow \cite{Scott2010} by assuming that the average number of factors in the model is fixed, irrespective of the number of factors available for selection, $P$. This prior has the effect of  decreasing $w_j$ as the number of factors under consideration increases, and thus accounts for multiplicity.

To  gauge the sensitivity of inference to this subjective choice of prior and to assess the importance of a factor to an individual's NEET status we perform sensitivity analysis on this prior value, see Figure \ref{fig:gamma_prior_sensitivity_plot} in the Results section for details.

The full model and prior specification is
\begin{eqnarray}
  y|\bm \beta,\bm \gamma &\sim& \mbox{Be} \left(\Phi(X_{A_1}\beta_{A_1})\right)\nonumber \\
  \beta_{A_1}|\bm \gamma& \sim &N\left(0,\Sigma_{A_1}\right) \nonumber \\
\beta_{A_0}|\bm \gamma& \sim & \delta(0)\nonumber \\
\bm \gamma &\sim& \prod_{k=1}^P \mbox{Be}(w_k).
 \label{eqn:model}\end{eqnarray}
where the notation $\mbox{Be}(p)$ denotes the Bernoulli distribution, with parameter $p$. 

\subsection*{Prediction and Estimation}
Our article takes a Bayesian approach and makes inference regarding the importance of factors via the joint posterior distribution $\Pr(\bm\gamma,\bm\beta|\bm y,X)$ and predicts the unobserved NEET status of an individual, denoted by $y^*$, given specific factor measurements, $x^*$, via its posterior predictive density, $\Pr(y^*=1|\mathbf y,X, x^*)$.  This density also accounts for the uncertainty surrounding the selection of factors, and the uncertainty surrounding the values of the regression co-efficients corresponding to a particular factor selection, by averaging over all possible factor combinations and values of the regression coefficients, where the average is w.r.t the joint posterior distribution of $\bm\gamma$ and $\bm\beta$. This is commonly referred to as marginalization. Specifically we compute
\begin{eqnarray}
\Pr(y^*\!\!=\!1|\mathbf y,\mathbf x, x^*)&=&\sum_{j=1}^{2^P}\int \Pr(y^*\!\!=\!1|\bm y,X, x^*,\bm\gamma_j,\bm\beta)p(\bm\beta| \mathbf y,X,\bm\gamma_j)d\bm\beta \;\Pr(\bm\gamma_j|\mathbf y,X)\nonumber\\
&\approx&\frac{1}{M}\sum_{m=1}^M \Pr(y^*\!\!=\!1|\mathbf y,X, x^*,\bm\gamma^{[m]},\bm\beta^{[m]})
\label{eq_marg}
\end{eqnarray}
where $\bm\gamma^{[m]}$ and $\bm\beta^{[m]}$ are samples drawn from the joint posterior $\Pr(\bm\gamma,\bm\beta|\mathbf y,X)$.  We approximate the LHS of equation \eqref{eq_marg}  by obtaining draw of $\bm \gamma^{[k]}_j$ and $\bm\beta^{[k]}_{\bm \gamma_j}$ using RJMCMC as in  \cite{Lamnisos2009}.  
The MCMC sampling scheme and diagnostics respectively appear in Supplementary Note 2. MCMC sampling scheme and Supplementary Note 3. Diagnostics.
The sampling scheme is coded in Matlab and can be downloaded from the Github repository https://github.com/divad-nhok/bayesian-variable-selection.

\subsection*{Data}
\cite{Purcell2015} describe the study methodology involving a longitudinal cohort design implemented within four HeadSpace youth mental health services in Australia.
We use this  dataset to compare the results from our approach with that of \cite{ODea2014}. 

\begin{figure}[htbp]
\center
\includegraphics[width=1.0\textwidth]{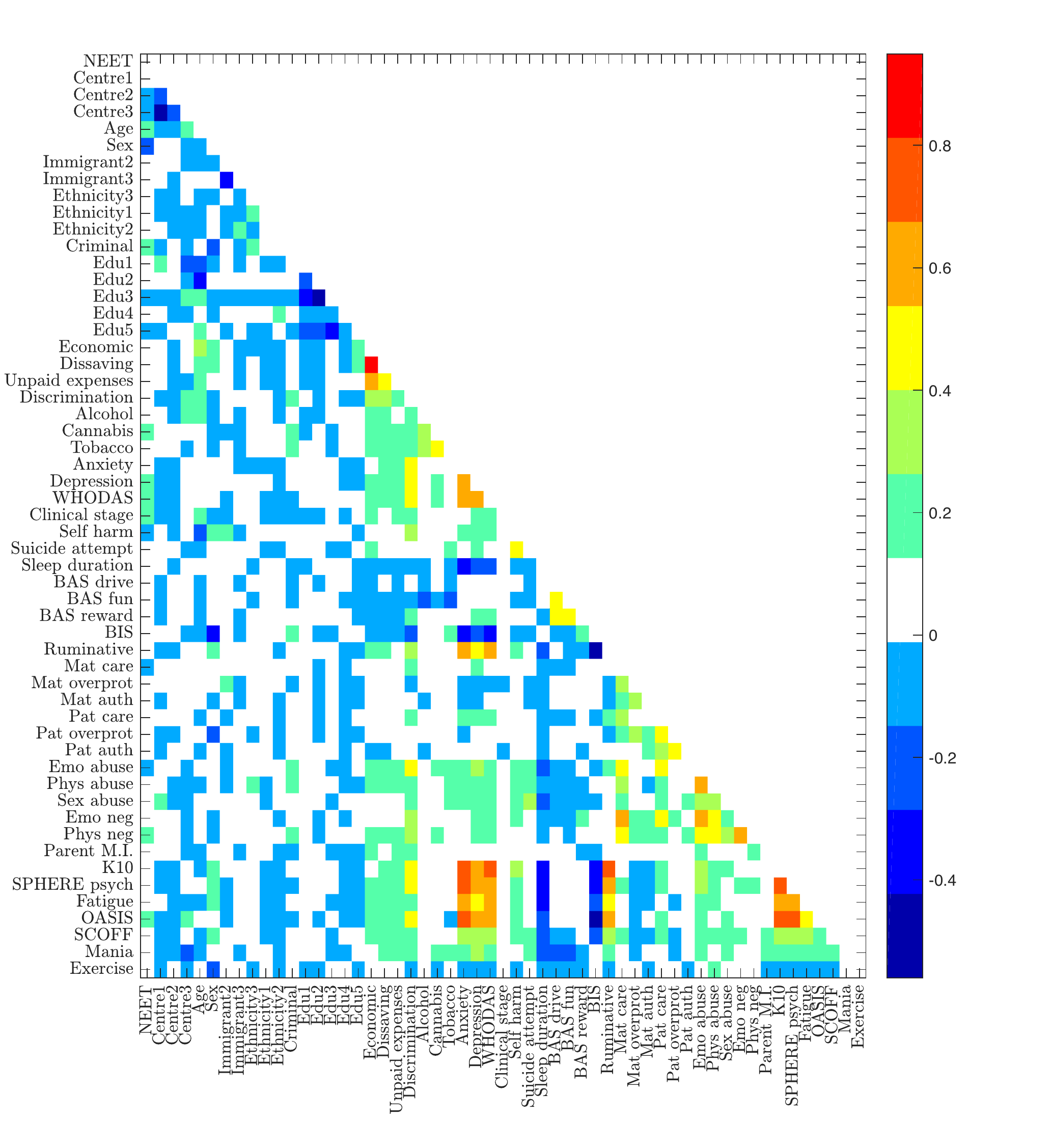}
\caption{
\textbf{Heatmap of correlations of factors for the \dslarge{}.}
Each cell corresponds to the correlation of two given factors. This figure shows that potential predictive factors are often highly correlated with other factors. For example the factor {\it Anxiety}, is strongly correlated with the factors {\it K10, Sphere Psych, Fatigue} and {\it OASIS}.  These high correlations suggest that confounds may be an issue.
}
\label{fig:corr_heatmap}
\end{figure}

The model selection procedure of \cite{ODea2014} is a two step process; single factor logistic regression models are estimated for those factors thought to be significant either from prior research or from expert opinion. 
Then a subset of these factors, those with single factor regression coefficients which are significantly different from  zero, as judged by p-value less than some pre-specified value, are included in a multiple regression where stepwise regression is run to select the best combination of factors.  
 
Although this is common practice, it is an ad-hoc procedure which can result in the omission of important factors in the final model, particularly in observational studies where confounds are problematic. 
A heatmap of  correlations between all possible factors in \cite{ODea2014} appears in Figure~\ref{fig:corr_heatmap}. Figure~\ref{fig:corr_heatmap} shows that about half of the pairwise correlations exceed, in absolute terms, the value of 0.08 (the approximate upper limit of a confidence interval for a sample correlation co-efficient in this context). For example the pairwise correlations between {\it Anxiety} and other factors exceeds 0.08 approximately 60\% of the time. The same is true for the {\it Rumination}. These values suggest that confounds may be an issue in this study. 

Additionally the stepwise procedure ignores model uncertainty, while 
in a Bayesian context  factor and model uncertainty are quantified by estimating the marginal  posterior probability ($MPP$),  and  joint posterior probability ($JPP$),  respectively of that model. The $MPP$ is the probability that a single factor is included in the model, after accounting for uncertainty in the selection of all other factors, and is denoted by $\Pr(\gamma_k=1|\bm y,\bm x)$ for $k=1,\ldots,P$. The $JPP$ is the probability that a given {\it set} of factors are included $\Pr(\bm\gamma|\bm y, \bm x)$, and prediction proceeds by weighting all possible model predictions by this probability as in equation \eqref{eq_marg}.  

There is an extensive statistical and bio-statistical literature on Bayesian variable selection and model averaging, which documents the advantages of Bayesian model averaging  over model selection procedures, such as stepwise regression, not only in terms of better predictions, but also in terms of more accurate statistical inference, (\cite{Raftery1997}, \cite{stodden2006model}, \cite{Wang04}, \cite{Viallefont01}, and \cite{Genell10}). 

The factors used by \cite{ODea2014}, together with the p-values for the single factor  logistic regression models in the first step, and again for the multiple logistic regression model, in the second step, appear in the last two columns of Table \ref{table:mpp}. 
We use shortened, descriptive factor names but the full names of the factors corresponding to factors listed in \cite{Purcell2015}  can be found in Supplementary Note 1. Factor names.

\subsection*{Data availability}
The datasets analysed during the current study are available at the Github repository https://github.com/divad-nhok/bayesian-variable-selection. Please note that the data associated with the factor {\it Ethnicity} have been omitted from this file but not from the analysis. This was done so that individuals could not be identified.

\section*{Results}
\label{sec:results}
The results section is split into four parts. 
First, we report the importance of single factors, as measured by the p-value for the \cite{ODea2014} analysis, and as measured by the $MPP$, for the Bayesian analysis. In addition we report the estimation and associated uncertainty of regression coefficients corresponding to these factors. 
Second, we report the importance of  factor combinations, as measured by the $JPP$.  
Third, the evaluation of predictive performance of various models is documented. 
Fourth, we represent results on the sensitivity of $MPP$ to changes in prior specifications. 

In each part we discuss three sets of results: 
the original results using a logistic regression,  based on a reduced set of factors, from \cite{ODea2014}, which we call the \dsoriginal{}; the results from the same factor set but using a Bayesian model, which we call the \dssmall{};  and the results from a model based on  all 50 potential predictors in the dataset as possible predictors, which we call the \dslarge{}.  

\subsection*{Identification of single factors}
\label{sec:results:mpp}
The single factor results appear in Table \ref{table:mpp}.
Columns 1 and 2, titled  $p_{\text{single}}$ and $p_{\text{multi}}$, contain the results reported by \cite{ODea2014}. 
Column 1 reports the p-value from running logistic regressions on single factors and column 2 reports the p-values of the regression coefficients from running a multiple logistic regression, containing the factors for which the p-value in the single factor models was less than $0.003$.

Columns 3 and 4, titled $\hat{\beta}_{\gamma=1}$, and $\hat{\sigma}_{\gamma=1}$, contain an estimate of the posterior mean and standard deviation of the regression coefficients respectively, conditional on that coefficient being non-zero. Column 5, titled $MPP$, is an estimate  of the marginal posterior probability of the factors considered in \cite{ODea2014}. 
These three  statistics, $\hat{\beta}_{\gamma=1}, \hat{\sigma}_{\gamma=1}$, and the $MPP$, were used to summarize the posterior distributions of the regression coefficients because these posterior distributions all consist of a discrete and continuous component, and the posterior mean and standard deviation, are not sufficient statistics to describe such distributions.

Columns 6, 7 and 8 contain the corresponding values for the \dslarge{}. 
Note that the results corresponding to the \dslarge{} do not contain the results for all factors included in the \dslarge{}, but instead show the same 20 factors used in the \dssmall{} as well as those factors with the next ten highest $MPP$s, after those factors already included in the \dssmall{}.
The full results for the \dslarge{} can be found in Supplementary Note 4. Results. 
The $MPP$s for individual factors are directly interpretable as the probability a factor is useful in predicting an individual's NEET status, after accounting for the uncertainty in all other factors. 

\begin{table}[!htbp]
\centering
\ra{1.3}
\scriptsize
\begin{tabular}{l l l l l l l l l l l}
\toprule
& \multicolumn{2}{c}{ \preoriginal } & \multicolumn{3}{c}{ \presmall } & \multicolumn{3}{c}{ \prelarge } \\ \cmidrule{2-3}  \cmidrule{4-7} \cmidrule{8-11}                                                                                  
Variable & {$p_{\text{single}}$} & {$p_{\text{multi}}$} & {$\hat{\beta}_{\gamma=1}$} & {$\hat{\sigma}_{\gamma=1}$} & {MPP \%}  & {$\hat{\beta}_{\gamma=1}$} & {$\hat{\sigma}_{\gamma=1}$} & {MPP \%} \\                        
\midrule                                                                                                                                                                                                                        
Criminal & 0.00 & 0.02 & 0.65 & 0.24 & 26 & 0.70 & 0.24 & 25 \\                                                                                                                                                   
Clinical stage & 0.00 &  & 0.56 & 0.21 & 21 & 0.57 & 0.21 & 17 \\                                                                                                                                                
Economic & 0.00 & 0.08 & 0.33 & 0.18 & 4 & 0.36 & 0.20 & 3 \\                                                                                                                                                    
Sex & 0.00 & 0.01 & -0.55 & 0.16 & 68 & -0.57 & 0.17 & 60 \\                                                                                                                                                     
Functioning & 0.00 &  & 0.20 & 0.10 & 3 & 0.24 & 0.11 & 3 \\                                                                                                                                                     
Depression & 0.00 & 0.00 & 0.32 & 0.09 & 85 & 0.35 & 0.10 & 86 \\                                                                                                                                                  
Age & 0.00 & 0.00 & 0.28 & 0.08 & 59 & 0.29 & 0.08 & 55 \\                                                                                                                                                        
Cannabis & 0.00 & 0.08 & 0.15 & 0.07 & 2 & 0.15 & 0.07 & 2 \\                                                                                                                                                  
Anxiety & 0.01 &  & -0.04 & 0.12 & 1 & -0.01 & 0.13 & 0 \\                                                                                                                                                       
Discrimination & 0.02 &  & 0.11 & 0.11 & 1 & 0.11 & 0.13 & 0 \\                                                                                                                                                  
Tobacco & 0.05 &  & 0.07 & 0.09 & 0 & 0.09 & 0.10 & 0 \\                                                                                                                                                         
Alcohol & 0.19 &  & -0.12 & 0.09 & 1 & -0.11 & 0.09 & 1 \\                                                                                                                                                      
Tertiary edu & 0.50 &  & -0.44 & 0.24 & 5 &  &  &  \\                                                                                                                                                        
Immigrant dichotomous & 0.65 &  &  &  &  &  &  &  &  &  \\                                                                                                                                                                      
Indigenous & 0.75 &  & 0.14 & 0.38 & 2 &  &  &  \\                                                                                                                                                                    
Centre & 0.92 &  &  &  &  &  &  &  &  &  \\                                                                                                                                                                                     
Ethnicity1 &  &  &  &  &  & 6.31 & 6.31 & 32 \\                                                                                                                                         
Education1 &  &  &  &  &  & 0.48 & 0.29 & 3 \\                                                                                                                                          
Education2 &  &  &  &  &  & 0.47 & 0.22 & 5 \\                                                                                                                                                                        
Dissaving &  &  &  &  &  & 0.38 & 0.4 & 24 \\                                                                                                                                                                         
Parent M.I. &  &  &  &  &  & 0.38 & 0.18 & 4 \\                                                                                                                               
Dissaving &  &  &  &  &  & 0.38 & 0.21 & 3 \\                                                                                                                                           
Unpaid expenses &  &  &  &  &  & 0.31 & 0.20 & 2 \\                                                                                                                                     
Phys neg &  &  &  &  &  & 0.26 & 0.10 & 5 \\                                                                                                                                           
BAS fun &  &  &  &  &  & 0.25 & 0.09 & 5 \\                                                                                                                                             
Mat auth &  &  &  &  &  & 0.22 & 0.08 & 12 \\                                                                                                                                        
BIS &  &  &  &  &  & 0.21 & 0.09 & 3 \\                                                                                                                                                 
OASIS &  &  &  &  &  & 0.18 & 0.12 & 1 \\                                                                                                                                               
Sleep time &  &  &  &  &  & 0.17 & 0.08 & 2 \\                                                                                                                                          
Sexual abuse &  &  &  &  &  & 0.16 & 0.10 & 1 \\                                                                                                                                        
Pat overprot &  &  &  &  &  & 0.15 & 0.09 & 1 \\                                                                                                                                                                   
\bottomrule                                                                                                                                                                            
\end{tabular}                                                                                                                                                                           
\caption{
\textbf{Single factor results.}
\preoriginal{}: $\text{p}_{\text{single}}$ and $\text{p}_{\text{multi}}$ show the p-values for the regression coefficients for the factors in the single and multiple factor logistic regressions respectively. 
\presmall{}: shows the results for the Bayesian analysis of the \dssmall{}. 
$\hat{\beta}_{\gamma=1}$ and $\hat{\sigma}_{\gamma=1}$ are estimates of posterior means and standard deviations respectively of the regression coefficients, conditional on the regression coefficient being non-zero. \prelarge{}: shows the results for the Bayesian analysis of the \dslarge{}, although we note that only the factors corresponding to those in the \dssmall{}, as well as the next 10 highest ranked $MPP$ factors are displayed. 
Column labels are the same as for \presmall{}. 
See Supplementary Note 1. Factor names for further information on the factors. 
}
\label{table:mpp}                                                                                                                                                                     
\end{table}  

\subsection*{Identification of factor combinations}
\label{sec:results:jpp}
Table \ref{table:jpp_small} summarizes the results for the top 20 combinations of factors (as measured by the $JPP$) for the \dssmall{}. 
The first column lists the factor combinations while the second column contains the joint posterior probability, $JPP$, of those factors. 
Column 3 contains the area-under-the-curve (AUC) of the Receiver Operating Characteristic (ROC) curve for predictions obtained by running a cross-validated logistic regression using those factors in column 1. Table \ref{table:jpp_large} reports analogous results for the \dslarge{}. 
 
\begin{figure}[htbp]
\center
\includegraphics[width=0.8\textwidth,height=0.8\textheight]{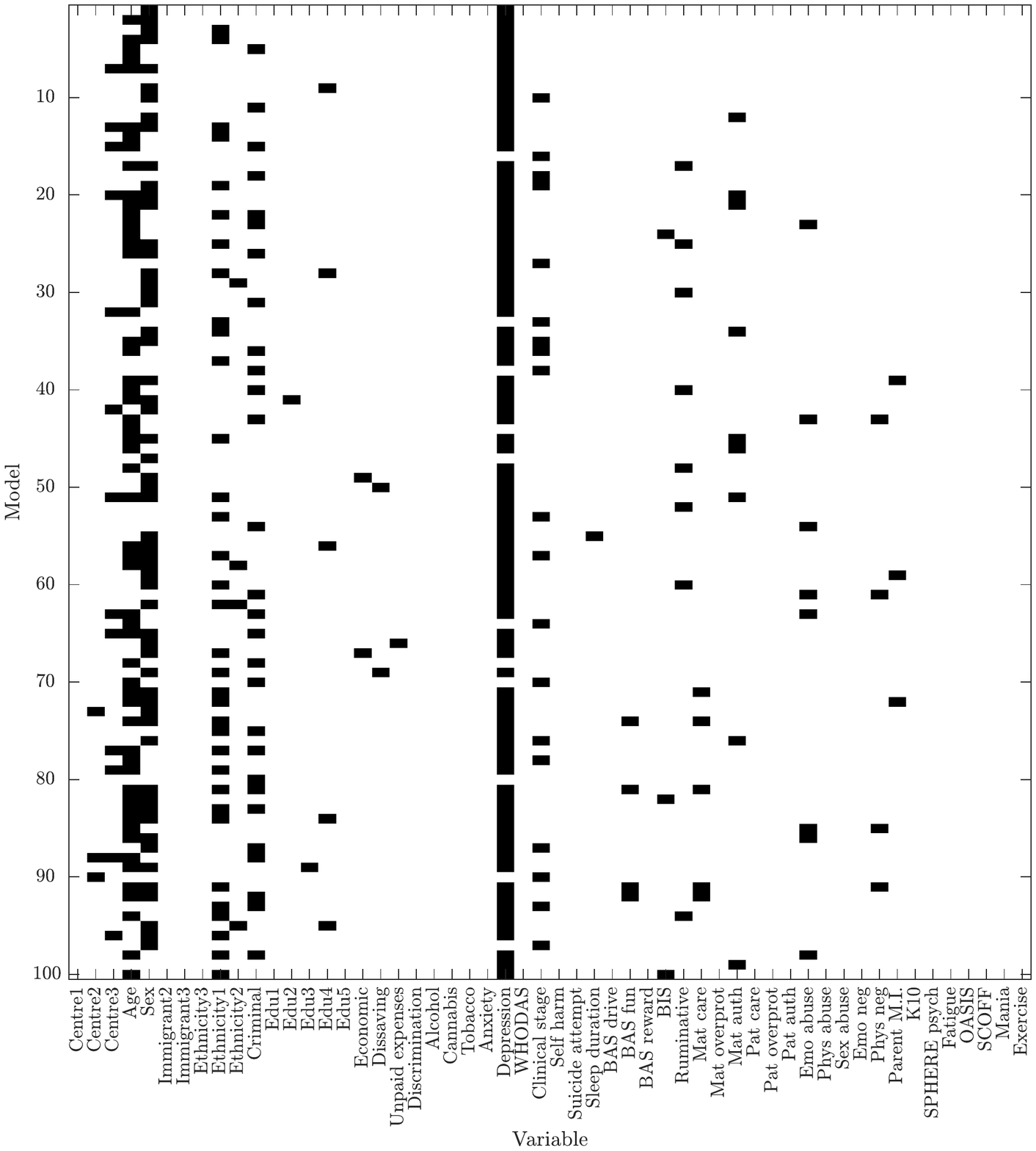}
\caption{
\textbf{$JPP$ heatmap for \dslarge{}.}
The heatmap displays the factors in the top 100 models for the \dslarge{} sorted by their descending $JPP$s. 
Each row corresponds to a model and each column corresponds to a given factor being in that model.
A black cell represents a given factor being included in a given model. 
}
\label{fig:jpp_heatmap}
\end{figure}

Figure \ref{fig:jpp_heatmap} contains a visualization to summarize how factors contribute to each of the models.
Each row corresponds to a model and each column corresponds to a given factor being in that model.
A black cell represents a given factor being included in a given model. 
The inference is that a column which is almost completely black indicates that this variable is present in the majority of the models and a row which is nearly black corresponds to a model which contains all variables.  

\begin{table}[!htbp]                                                                    
\centering                                                                              
\ra{1.3}                                                                                
\scriptsize                                                                             
\begin{tabular}{@{}l l l l@{}}                                                        
\toprule                                                                                             
{\it Factor combinations} & {$JPP$ \%} & {AUC\%} \\
\midrule                                                                                
Age, Sex, Depression & 15 & 70  \\         
Sex, Depression & 14 & 66  \\                                                  
Centre3, Age, Sex, Depression & 7 & 71 \\                                     
Age, Criminal, Depression & 4 & 68  \\                                         
Age, Depression & 3 & 67  \\                                                   
Sex, Depression, Clinical stage & 3 & 67  \\                                   
Age, Sex, Criminal, Depression & 2 & 70 \\                                    
Centre3, Age, Criminal, Depression & 2 & 71  \\                                
Age, Sex, Depression, Clinical stage & 2 & 70 \\                                                            
Depression & 2 & 63  \\ 
Sex, Criminal, Depression & 2 & 67  \\ 
Centre3, Sex, Depression & 1 & 68  \\               
Centre3, Age, Sex, Criminal, Depression & 1 & 72  \\          
Criminal, Depression, Clinical stage & 1 & 67  \\                              
Criminal, Depression & 1 & 65  \\                                              
Centre3, Age, Depression & 1 & 69  \\                                          
Age, Criminal, Depression, Clinical stage & 1 & 70  \\                         
Depression, Clinical stage & 1 & 65 \\                                        
Sex, Economic, Depression & 1 & 67  \\                                         
Clinical stage & 1 & 52  \\                                                   
\bottomrule                                                                             
\end{tabular}                                                                           
\caption{
\textbf{\dssmall{} Results for Combinations of factors.}
This table contains the factor combinations with the 20 highest $JPP$s. 
The column {\it factor combinations} contains the factors in a given model. 
$JPP$ contains the joint posterior probability, expressed as a percentage, for that model. 
The column headed AUC contains the ROC area-under-curve percentage
 derived by running a separate logistic regression cross-validation using the selected factors from the Bayesian model.}  
\label{table:jpp_small}                                                                   
\end{table}

\begin{table}[!htbp]                                                                                                                                                               
\centering                                                                                                                                                                         
\ra{1.3}                                                                                                                                                                           
\scriptsize                                                                                                                                                                        
\begin{tabular}{@{}l l l l l@{}}                                                                                                                                                   
\toprule                                                                                                                                                                           
                                                                                                         
Factor combinations & {JPP \% $\times 10^{-4}$} & {AUC \%} \\                                                                          
\midrule                                                                                                                                                                           
Sex, Depression & 5 & 66 \\                                                   
Age, Sex, Depression & 4 & 69  \\                                              
Sex, Ethnicity1, Depression & 2 & 66 \\                                       
Age, Sex, Ethnicity1, Depression & 2 & 69 \\                                  
Age, Criminal, Depression & 2 & 69  \\                                         
Age, Depression & 1 & 68 \\                                                   
Centre3, Age, Sex, Depression & 1 & 71 \\                                     
Depression & 1 & 63 \\                                                        
Sex, Education4, Depression & 1 & 66 \\                                       
Sex, Depression, Clinical stage & 1 & 68 \\                                   
Criminal, Depression & 1 & 64 \\                                              
Sex, Depression, M.authority & 1 & 68 \\                                      
Centre3, Age, Sex, Ethnicity1, Depression & 1 & 72 \\                         
Age, Ethnicity1, Depression & 1 & 68 \\                                       
Centre3, Age, Criminal, Depression & 1 & 70 \\                                
Clinical stage & 1 & 54 \\                                                    
Age, Sex, Depression, Ruminative & 1 & 70 \\                                  
Criminal, Depression, Clinical stage & 1 & 67 \\                              
Sex, Ethnicity1, Depression, Clinical stage & 1 & 66 \\                       
Centre3, Age, Sex, Depression, M.authority & 1 & 72 \\                         
\bottomrule                                                                                                                                                                        
\end{tabular}                                                                                                                                                                      
\caption{
{\bf $JPP$s and AUC for factor combinations using \dslarge{}}.
This table contains the factor combinations with the 20 highest $JPP$s. 
The column {\it factor combinations} contains the factors in a given model. 
The $JPP$ is joint posterior probability, expressed as a percentage, for that model and AUC contains the ROC area-under-curve percentage for  the logistic regression AUC, derived by running a separate logistic regression cross-validation using the selected factors from the Bayesian model.}
\label{table:jpp_large}                                                                                                                                                                     
\end{table}

\subsection*{Evaluation of predictive performance}
\label{sec:results:models}
In this section we evaluate the predictive performance of models, measured by the area-under-the-curve (AUC) of the (ROC) curve using 5 fold cross validation.   The factor combinations evaluated are produced for both the \dssmall{} and the \dslarge{}, using two methods.   

The first method is to order models by their $JPP$ and the predictive results for models formed in this way are reported in column 3 of Tables \ref{table:jpp_small} and \ref{table:jpp_large}.
The second method is to order models by sequentially adding factors in order of their $MPP$.   
For example for the  \dslarge{} the first model would contain the factor \textit{Depression}, while the second would contain \textit{Depression} and \textit{Sex}, and  the third, \textit{Depression}, \textit{Sex} and \textit{Age}, etc. 
The corresponding models for the  \dssmall{} are formed similarly.

\begin{figure}[htbp]
\center
\includegraphics[width=1.0\textwidth]{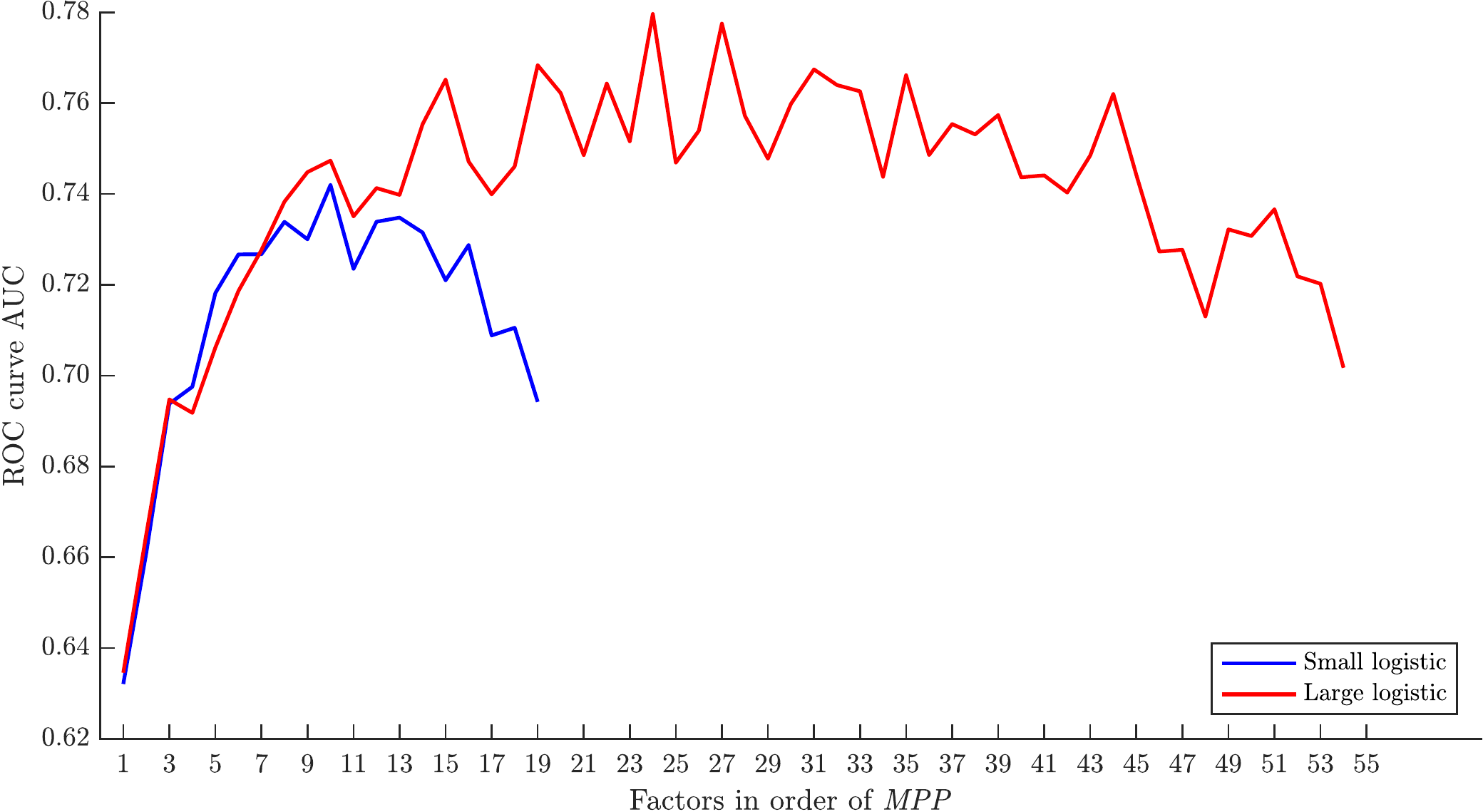}
\caption{
\textbf{AUCs for nested models.}
Each curve is constructed by first ranking the MPPs of each factor for a given dataset from highest to lowest. 
NEET status is then predicted using the highest ranked factors, until we have created a model that includes all factors. 
}
\label{fig:nested_auc}
\end{figure}

Figure \ref{fig:nested_auc} plots the predictive performance for the models formed in order of the $MPP$, for both the \dssmall{} (blue) and for  \dslarge{} (black) and shows that those factors with the highest MPP contribute to the greatest marginal increases in the AUC, while adding subsequent factors has little or negative effect when added to the model, as evidenced in the relatively flat curves for the weighted models after approximately the first five factors. 

\subsection*{Using priors to assess factor importance}
\label{sec:results:priors}

\begin{figure}[htbp]
\center
\includegraphics[width=.8\textwidth]{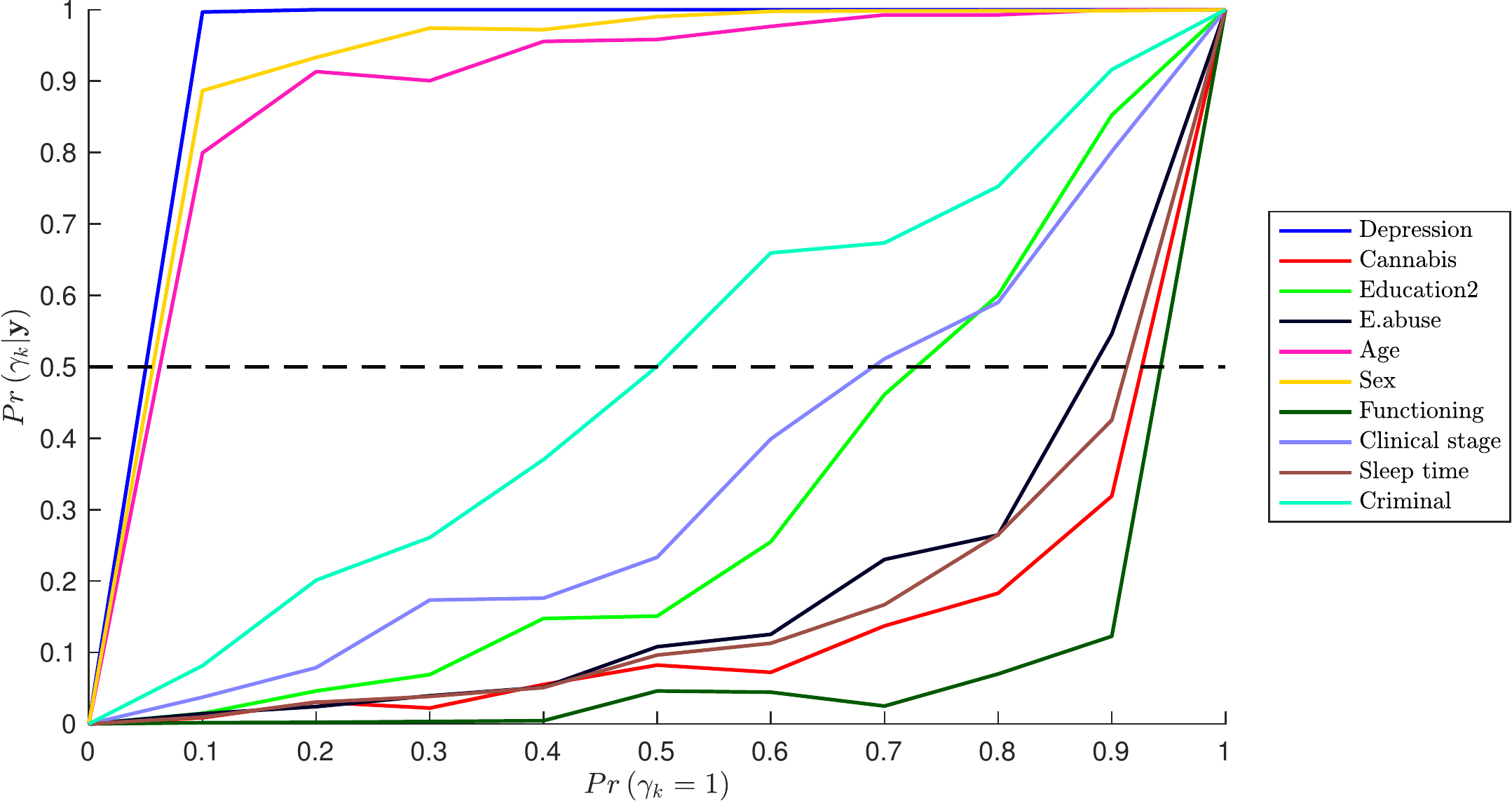}
\caption{
\textbf{Sensitivity analysis for the prior.}
The effect of  $\Pr \left(\gamma_k = 1 \right)$ on $\Pr \left(\gamma_k = 1|\bm y,\bm x \right)$,
i.e. how the $MPP$ changes as we change the prior probability of inclusion for a given factor. All the remaining factors for which the prior probability, $w$, did not vary, had this value fixed at 0.78.  This number was chosen to maximize the AUC for the ROC curve.
}
\label{fig:gamma_prior_sensitivity_plot}
\end{figure}

Figure \ref{fig:gamma_prior_sensitivity_plot} shows the sensitivity of posterior inference to the prior probability of factor inclusion. 
The prior probability that a factor is included is given by $\Pr(\gamma_k=1)=w_k$, for $k=1,\ldots,P$, as in equation \ref{eqn:model}. 
Inference regarding that factor's inclusion is via the marginal posterior distribution $\Pr(\gamma_k=1|\bm y,\bm x)$. Figure \ref{fig:gamma_prior_sensitivity_plot} shows that
\textit{Depression}, \textit{Sex} and \textit{Age} are relatively insensitive to changes in the prior probability and appear important factors regardless of the prior.
Similarly, altering the prior probability of inclusion of \textit{WHODAS}, \textit{Cannabis}, \textit{Sleep Duration} and \textit{Emo Abuse} has little effect on the posterior probability; these factors are unlikely to be important irrespective of the prior. 
In contrast, inference regarding factors such as \textit{Criminal}, \textit{Clinical Stage} and \textit{Education} influence depends very much on how much weight is given to their prior probability of inclusion. 

\section*{Discussion}

In this section we demonstrate a number of issues that contribute to difficulties in inferences in epidemiological data, and subsequently to the irreproducibility of results, by comparing the results of our Bayesian analysis, with that of  \cite{ODea2014}.

\subsection*{Impact of data choice on single factor inference} 
The first issue we address is the impact  of the choice of factors to include and data preprocessing on the analysis.  
In situations where the number of factors $P$ is relatively large it is common,  and sometimes desirable, to reduce the number of factors available for prediction, using expert knowledge  or results of prior studies.    
\cite{ODea2014} did just this and the number of factors considered for prediction was reduced from approximately 50 to 20.  
The decision to omit factor $k$ from a model is equivalent to stating that {\it a priori}, the probability of inclusion is identically zero, i.e. $\Pr(\gamma_k=1)=0$.  
This is a very strong prior and could be replaced with  $\Pr(\gamma_k=1)=w_k$, and letting $w_k$ vary, as shown in Figure \ref{fig:gamma_prior_sensitivity_plot}.

The last column in Table \ref{table:mpp} shows the $MPP$ for the most probable factors when the entire dataset is used and highlights that the decision to exclude variables can result in missing many important factors, such {\it Centre Location}, \textit{Ethnicity1}, \textit{Clinical Stage}, \textit{Dissavings} and \textit{Mat Auth}, to name a few.  

\begin{figure}[htbp]
\center
\includegraphics[width=1.0\textwidth]{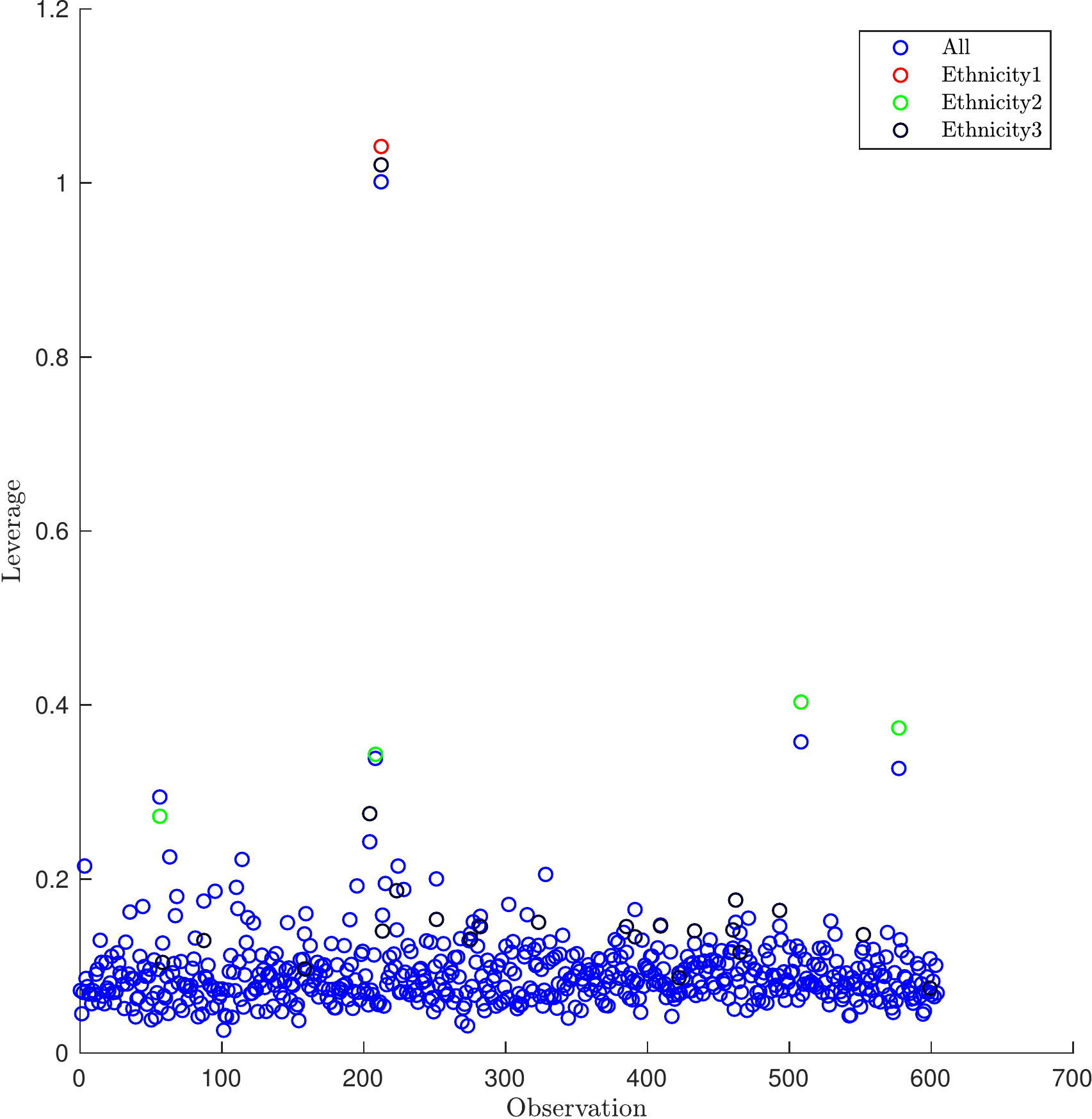}
\caption{
\textbf{Observations vs. leverage}
Points have been jittered to make them distinct. Blue points are the leverage for all  observations. Red green and black points are leverage values corresponding to observations for individuals with \textit{Ethnicity1}, \textit{Ethnicity2} and  \textit{Ethnicity3} respectively.
}
\label{fig:leverage}
\end{figure}

Another aspect of data choice is to identify observations that may strongly influence inference drawn from the study.  For example the $MPP$ for {\it Ethnicity1}, is equal to 0.32.  There is only one individual, out of approximately 600, with this ethnicity.  The measurements on the factors associated with this individual have values which make this individual an outlier in the covariate space, meaning that the values for this individual have more than average influence over the fitted values. Known as high leverage points, these points pull the fitted values  towards these observations. 
Figure \ref{fig:leverage} shows a plot of the leverage values for all observations, and shows that the individual with {\it Ethnicity1} has  extremely  high leverage values.  
Clearly we would not want to base a policy decision, for example the development of an intervention strategy for this particular ethnic group, based on a single observation.

Furthermore  the posterior distribution of the regression coefficient for the  factor \textit{Ethnicity1}, shown in Supplementary Figure S10, is very non-normal, the majority of iterates of the regression coefficient, ($\approx 75\%$), are positive, a small proportion of iterates are negative, $\approx 11\%$, but large, (a consequence of related factors being in or out of the Bayesian model at each iteration), while $\approx 68\%$ are identically zero.  Clearly it is not meaningful to discuss the distribution of a regression coefficient which is based on a single observation, but it does highlight the importance of rigorous data analysis.

\subsection*{Impact of analytic technique on single factor inference}
The second issue we address is the impact of the choice of analytic technique, Bayesian vs frequentist, has on inference concerning the individual factors that affect an individual's likelihood of becoming NEET. 

The final model of \cite{ODea2014} selected \textit{Age} ($p=0.03$), \textit{Sex} ($p=0.05$), \textit{Criminal} ($p=0.02$) and \textit{Depression} ($p=0.00$) as significant at the $5\%$ level, while \textit{Cannabis} ($p=0.08$) and \textit{Economic} ($p=0.08$) were also significant at the $10\%$ level. 
This is  similar to the Bayesian inference for the \dssmall{}  with notable exceptions.  
The Bayesian analysis of the \dssmall{}, identifies \textit{Depression}, \textit{Age} and \textit{Sex} as the top three important factors; consistent with   \cite{ODea2014}. 
However the factor \textit{Clinical Stage}, identified as having the fifth highest $MPP$ in the Bayesian analysis, had a p-value of $>0.1$, in the final model in \cite{ODea2014}, and was therefore deemed insignificant. 
Conversely factors such as \textit{Cannabis} and \textit{Economic}, while significant at the 10\% levels in the \cite{ODea2014} analysis, had very low $MPP$'s in the Bayesian analysis.

The difference in inference between the Bayesian analysis of the \dslarge{} and the  \cite{ODea2014} analysis, is even more marked. 
Now factors such as \textit{Clinical Stage},    \textit{Mat Auth}, and \textit{Dissaving} emerge as important factors. 
These factors were either discarded in the modelling process or not considered at all in  \cite{ODea2014}.

\subsection*{Impact of analytic technique on inference for combinations of factors}
For our discussion we categorize predictive factors into those which are non-modifiable, such as \textit{Age} and \textit{Sex}, and those which are modifiable i.e. have the potential to be changed by some intervention. 
For the \dsoriginal{} and \dssmall{} inference regarding the non-modifiable factors,  \textit{Age} and \textit{Sex}, is consistent; for the \dssmall{}  dataset  \textit{Age} and \textit{Sex} are in 70\% and 60\% of the top 10 models respectively while for the \dslarge{} the corresponding figures are 50\% and 70\%. 
However inference regarding  the modifiable factors is mixed.  
For example, Depression is in 95\% of the top 20 factors combinations for the \dssmall{} and \dslarge{}.  \textit{Clinical Stage}, is in 25\% of the top 20 models in the \dssmall{} and in 20\% of the top 20 models in the \dslarge{} but not at all in the frequentist analysis while \textit{Cannabis} is identified as important only in the frequentist analysis.  

Note that the Bayesian analyses of both datasets identify one of the \textit{headspace} \textit{Centre} locations, \textit{Centre3} as an important factor, whereas this factor, as a dichotomous variable indicating the centre being in Sydney or Melbourne, did not even make it past the initial variable selection procedure in the \cite{ODea2014} analysis.  
We suspect that this is because of data preprocessing and the arbitrary manner in which the variable \textit{Centre} was coded, and this is the subject of future research. 
Our results suggest that \textit{Centre3} is a proxy for other factors, not measured, associated with the location of that centre, and perhaps offers an avenue of future research to determine what those others factors might be, an avenue that would not be apparent in the previous analysis of \cite{ODea2014}.

With respect to inference, Table \ref{table:jpp_large} shows that, in addition to the core group of non-modifiable factors, \textit{Age} and \textit{Sex}, other non-modifiable factors such as Ethnicity (\textit{Ethnicity1, Ethnicity2, Ethnicity3}) become important.  
And for modifiable factors a whole sluice of additional factors appear to be related to NEET status, including  \textit{Emo Abuse}, \textit{Education}, \textit{Mat Auth}, \textit{Fatigue}, \textit{Parental M.I.} and \textit{Phys Neg}.  
These mixed results for modifiable factors clearly present problems when recommending intervention strategies.

\subsection*{Predictive performance}
The results in Tables \ref{table:jpp_small} and  \ref{table:jpp_large} show just how difficult selecting the one best model can be; multiple models have similar predictive performance yet very different factor combinations. 
For example, if we were to introduce interventions based on the most probable model (highest $JPP$) then we would target \textit{Depression} only, while if those interventions were based on the model with the highest predictive ability (largest AUC) then those we would target  \textit{Mat Auth} and  \textit{Centre3}, as well as \textit{Depression}.  

Furthermore many models have comparable predictive performance, as measured by AUC, yet different factor combinations.  If prediction of the outcome is the only aim, then this inconsistency in inference can be overlooked, but it presents a problem if the aim is to develop invention strategies to reduce youth disengagement.
\\
\\
\noindent In summary, given that public policy is often based on epidemiological associations rather than formal intervention studies, this lack of reproducibility and the impact of different research groups' prior assumptions, which are often implicit rather than explicit, really do matter.  However it is worth noting that {\it Depression}, a potentially modifiable factor, is robustly associated with youth disengagement across all models and inference regarding this factor is relatively independent of prior probabilities. Hence the study of how this factor impacts youth disengagement is worthy of support. For example  does depression cause disengagement or it is a consequence of it? What is the best treatment for depression and does this vary with age and/or NEET status? More specific intervention studies, which attempt to control for other sources of uncertainties, are needed to answer these questions.  In contrast factors such as {\it Clinical Staging} and {\it Education} need more investigation before being easily translated into larger scale policy changes.

\section*{Conclusion}

When dealing with observational data there are many sources of variability in the journey to discovery;  variability due to data selection and preprocessing, variability due to the type of analysis performed, variability due to the models proposed, and variability due to sampling.  Each source of variability builds on the previous one. 
Given that p-values were  designed to capture only the variability due to sampling, it is not surprising that results published on the basis of `statistical significance', as measured by p-values, are often irreproducible. 

In this paper we have  taken a relatively small dataset, and shown that while the predictive capabilities of different models, are similar, inference regarding which factors effect a young person's propensity to become disengaged, varies with the data used and the type of analysis performed. Even for a fixed set of data and choice of analysis, the Bayesian approach showed that models with very different factor combinations have similar predictive ability.  Therefore if models are selected on the basis of predictive performance, it is not surprising that research findings based on model selection techniques have a high level of irreproducibility.  
 
As the volume of data available for analysis increases, so too will the researcher's degrees of freedom and if we continue to ignore other sources of variability then this will lead to an increase in irreproducibility; nothing is more certain. What is needed is an acknowledgment and acceptance of the various sources of uncertainty, and a rethink of how we measure it, report it, and take it into account when making decisions.  

We argue strongly that, in addition to making data and code available, journal editors should require researchers to perform different types of analyses, using different data preprocessing techniques, and different models, such as presented in this paper.  It is in the comparison of differing sets of results, that insights into the issue at hand become more apparent. 

\bibliography{sample}

\end{document}